\documentclass
[12pt,prd,nofootinbib,eqsecnum,titlepage,preprint,groupedaddress,showpacs,showkeys]{revtex4}%
\usepackage{amsfonts}
\usepackage{amsmath}
\usepackage{amssymb}
\usepackage{graphicx}%
\begin{document}
\preprint{ }
\title{Real Scalar Fields on Manifolds}
\author{J.R. Morris}
\email{jmorris@iun.edu}
\affiliation{Physics Department, Indiana University Northwest, 3400 Broadway, Gary, Indiana
46408, USA}

\begin{abstract}
A generic theory of a single real scalar field is considered, and a simple
method is presented for obtaining a class of solutions to the equation of
motion. These solutions are obtained from a simpler equation of motion that is
generated by replacing a set of the original coordinates by a set of
generalized coordinates, which are harmonic functions in the spacetime. These
ansatz solutions solve the original equation of motion on manifolds that are
defined by simple constraints. These manifolds, and their dynamics, are
independent of the form of the scalar potential. Some scalar field solutions,
and manifolds upon which they exist, are presented for Klein-Gordon and
quartic potentials as examples. Solutions existing on leaves of a foliated
space may allow inferences of the characteristics expected of exact bulk solutions.

\end{abstract}
\maketitle

\section{Introduction}

Scalar fields play a prominent role in modern physical theories. Scalar
potentials with multiple vacuum states allow for the possible existence of
various types of topological and nontopological solitons, including kinks and
domain walls, cosmic strings, and magnetic poles\cite{Vilenkin,VSbook}. Scalar
field interactions can give rise to networks of
defects\cite{networks1,networks2} and nested defects\cite{nested}, where one
defect may form inside another (host) defect. Solitonic structures associated
with scalar moduli are found in dilatonic and low energy string
theories\cite{moduli}. The many interesting types of scalar field phenomena
serve to motivate the study of various kinds of scalar field theories and
their solutions. Often, attention is focused on a simplified scenario where
scalar fields depend upon only one or two coordinates, and solutions are
easier to obtain and analyze\cite{soliton}. Solutions to the equations of
motion that depend on several variables are generally less accessible, but may
contain a relatively rich structure.

Here, we present a simple ansatz allowing one to map a solution of fewer
coordinate variables to one of more coordinate variables. These ansatz
solutions, however, are subject to a caveat, in that they solve the equation
of motion only on a well defined manifold, or set of manifolds, in the
spacetime. The manifold(s) may consist of the entire spacetime, or may be in
the form of hypersurfaces within the spacetime. For a space that is foliated
by a set of surfaces, it seems natural to expect that the set of solutions on
the various leaves of the foliation will give an indication of the
mathematical and physical natures of an exact solution solving the equation of
motion in the spacetime bulk. This may provide a way to extract information
about complicated solutions of a scalar field theory that would be otherwise
hard to obtain.

We consider a theory of a single real scalar field described by an action%
\begin{equation}
S=\int d^{N}x\sqrt{g}\left[  \frac{1}{2}\partial_{\mu}\phi\partial^{\mu}%
\phi-V(\phi)\right]  \label{1}%
\end{equation}

in an $N=D+1$ dimensional spacetime with $D$ spatial dimensions, and
$\mu=0,\cdot\cdot\cdot,D$. A mostly negative metric is used with $g_{\mu\nu
}=(+,-,-,\cdot\cdot\cdot,-)$ and $g=|\det g_{\mu\nu}|$. The metric $g_{\mu\nu
}(x)$ is taken to be a nondynamical background field, and, for simplicity, we
take fields and coordinates to be dimensionless. The equation of motion (EoM)
is%
\begin{equation}
\square\phi=\nabla_{\mu}\partial^{\mu}\phi=-\frac{\partial V}{\partial\phi
}=-V^{\prime}(\phi) \label{2}%
\end{equation}

This 2nd order DE can be difficult to solve, especially if there is a
complicated potential $V(\phi)$ or a solution is sought where $\phi$ depends
on more than one or two coordinate variables. We therefore consider a
simplifying ansatz that will generate solutions to the EoM, but the solutions
generated by the ansatz generally exist on some set of manifolds or
hypersurfaces. For some cases, the manifold is the full spacetime. In other
cases, a continuous set of hypersurfaces can foliate the spacetime, or a
dynamical set of surfaces may move through the space. These manifolds can
therefore span the spacetime in one way or another and thereby give some
indication of, at least qualitative features, that exact \textquotedblleft
bulk\textquotedblright\ solutions (which may be hard to obtain directly) of
the EoM may be expected to exhibit. These ansatz solutions form a subset of
the full solution spectrum for the theory.

The ansatz is based on the idea that the function $\phi(x^{\mu})$ can depend
on the coordinates $x^{\mu}$ through a set of linearly independent functions
$q^{\alpha}(x^{\mu})$, where the number of functions $q^{\alpha}$ is less than
or equal to the number of spacetime coordinates $x^{\mu}$. The $q^{\alpha}$
serve as generalized coordinates, and must satisfy certain constraint
conditions in order for $\phi\lbrack q^{\alpha}(x^{\mu})]$ to satisfy the
original EoM. These constraints, in turn, define some manifold of dimension
$\leq N$ on which the solutions exist. These constraints are associated with a
$q-$ space metric, which has components that become Minkowski-valued on the
solution manifold. In addition, the functions $q^{\alpha}$ must be harmonic in
the original spacetime, satisfying $\square q^{\alpha}(x^{\mu})=0$. For the
case where the $q^{\alpha}$ consists of just one spacelike generalized
coordinate, say $q^{1}=\xi(x^{\mu})$, with $\phi=\phi\lbrack\xi(x^{\mu})]$,
the ansatz considered here reduces to a BPS-like ansatz where the solution
$\phi(\xi)$ can be obtained directly from the potential function $V(\phi)$.
The solution manifolds and their associated dynamics are independent of the
form of the scalar field potential.

In the following sections we present the solution generating ansatz. Some
concrete examples of solutions of scalar field theories, and manifolds on
which they exist, are then presented. These include theories with potentials
for massless and massive Klein-Gordon fields, as well as $\phi^{4}$ theory. We
focus on 1d and 2d cases, where $\phi$ depends upon only one or two $q$
functions, respectively. For the 1d case the generalized coordinate can be
either a timelike or a spacelike one. For the 2d case there can be one
timelike and one spacelike function, or two that are spacelike. Static and
dynamical solutions are obtained describing configurations such as
Klein-Gordon fields, kinks, and domain ribbons on various manifolds.

\section{The ansatz}

The purpose of our simplifying ansatz is to obtain solutions to the EoM in
(\ref{2}) by considering $\phi(x^{\mu})$ to have a dependence on coordinates
$x^{\mu}$ only through a set of linearly independent generalized coordinate
functions $q^{\alpha}(x^{\mu})$, i.e., $\phi(x^{\mu})=\phi\lbrack q^{\alpha
}(x^{\mu})]$. The number of generalized coordinates $q^{\alpha}$ is less than,
or equal to, the number of spacetime coordinates $x^{\mu}$. In other words,
the $\alpha$ indices can take any set of the values of the $\mu$ indices,
where $\mu=0,1,2,\cdot\cdot\cdot,D$. We could choose $q^{\mu}=x^{\mu}$ for
some of the coordinates, but we will focus on the case where the number of
$q^{\alpha}\neq x^{\alpha}$ is less than the total number of spacetime
coordinates $\{x^{\mu}\}$, and therefore $\phi(q^{\alpha})$ is a function of
$M<N$ generalized coordinates $q^{\alpha}(x)\neq x^{\alpha}$.

Using a notation where differentiation with respect to a $q$ coordinate is
denoted by an overbar, $\bar{\partial}_{\alpha}=\partial/\partial q^{\alpha}$,
we write%
\begin{equation}
\partial_{\mu}\phi=\left(  \partial_{\mu}q^{\alpha}\right)  \bar{\partial
}_{\alpha}\phi,\ \ \ \ \ \partial^{\mu}\phi=\left(  \partial^{\mu}q^{\alpha
}\right)  \bar{\partial}_{\alpha}\phi,\ \ \ \ \bar{\partial}_{\alpha}%
\equiv\frac{\partial}{\partial q^{\alpha}} \label{3}%
\end{equation}
The term $\square\phi$ on the left hand side of (\ref{2}) can be written as%
\begin{equation}%
\begin{array}
[c]{ll}%
\square\phi & =\nabla_{\mu}\partial^{\mu}\phi=\dfrac{1}{\sqrt{g}}\partial
_{\mu}(\sqrt{g}\partial^{\mu}\phi)=\dfrac{1}{\sqrt{g}}\partial_{\mu}[\sqrt
{g}\left(  \partial^{\mu}q^{\alpha}\right)  \bar{\partial}_{\alpha}\phi]\\
& =(\square q^{\alpha})\bar{\partial}_{\alpha}\phi+(\partial_{\mu}q^{\alpha
}\partial^{\mu}q^{\beta})\bar{\partial}_{\alpha}\bar{\partial}_{\beta}\phi
\end{array}
\label{4}%
\end{equation}

The EoM of (\ref{2}) then takes the form%

\begin{equation}
\square\phi+V^{\prime}(\phi)=\left(  \square q^{\alpha}\right)  \bar{\partial
}_{\alpha}\phi+\left(  \partial_{\mu}q^{\alpha}\partial^{\mu}q^{\beta}\right)
\bar{\partial}_{\alpha}\bar{\partial}_{\beta}\phi+V^{\prime}(\phi)=0 \label{5}%
\end{equation}
\ 

We consider a class of solutions that satisfy the simplified EoM in the $q-$
space,
\begin{equation}
\eta^{\alpha\beta}\bar{\partial}_{\alpha}\bar{\partial}_{\beta}\phi+V^{\prime
}(\phi)=0 \label{6}%
\end{equation}

subject to the conditions%
\begin{equation}
\square q^{\alpha}=0,\ \ \ \ \partial_{\mu}q^{\alpha}\partial^{\mu}q^{\beta
}=\eta^{\alpha\beta} \label{7a}%
\end{equation}

The 1st condition requires $q^{\alpha}(x^{\mu})$ to be a harmonic function,
$\square q^{\alpha}=\nabla_{\mu}\partial^{\mu}q^{\alpha}=0$, and the 2nd
condition imposes a set of constraints upon the $q^{\alpha}$. This set of
constraints must be satisfied simultaneously. Each constraint equation can
lead to a constraint between the coordinates $x^{\mu}$, and can therefore
define a manifold. The solution manifold $\mathcal{M}$ is the intersection of
all of the individual constraint manifolds.

To summarize, we can generate a solution $\phi(x^{\mu})$ of the EoM by
considering a solution $\varphi(x^{\alpha})$ that solves an equation of motion
of the form $\eta^{\alpha\beta}\partial_{\alpha}\partial_{\beta}%
\varphi+V^{\prime}(\varphi)=0$ in a Minkowski spacetime or Euclidean space,
with $\varphi$ depending on a set of coordinates $x^{\alpha}$ that is a subset
of the spacetime coordinates $x^{\mu}.$ We then make replacements $x^{\alpha
}\rightarrow q^{\alpha}(x^{\mu})$ and $\varphi(x^{\alpha})\rightarrow
\phi\lbrack q^{\alpha}(x^{\mu})]$ to obtain the $q-$ space equation of motion
in (\ref{6}). This function $\phi(q^{\alpha})$ will also be a solution to the
original EoM in (\ref{2}) on the manifold $\mathcal{M}$, provided that the
conditions in (\ref{7a}) are satisfied. Each function $q^{\alpha}(x^{\mu})$ is
harmonic in the original spacetime, and the constraint equations
$\partial_{\mu}q^{\alpha}\partial^{\mu}q^{\beta}=\eta^{\alpha\beta}$ define
the solution manifold $\mathcal{M}$ where all constraints are satisfied
simultaneously. Then the EoM is satisfied on $\mathcal{M}$, i.e.,%
\begin{equation}
\left\{  \nabla_{\mu}\partial^{\mu}\phi+V^{\prime}(\phi)\right\}
\Big|_{\mathcal{M}}=0 \label{8}%
\end{equation}

Let us try to look at this in a slightly different way. Suppose that we have a
spacetime with $N$ coordinates $x^{\mu}$ and metric $g_{\mu\nu}(x)$. We then
define $N$ new generalized coordinates $q^{\mu}(x)$, although some of the
$q$'s may be identically equal to some of the $x$'s; e.g., $q^{m}=x^{m}$,
where $\{q^{m}\}$ is a proper subset of $\{q^{\mu}\}$. We then have nontrivial
functions $q^{\alpha}(x)$ for a subset $\{q^{\alpha}\}$ ($\alpha\neq m$). Now
consider a diffeomorphism that takes $x^{\mu}\rightarrow q^{\mu}$ and the
metric $g_{\mu\nu}(x)\rightarrow\bar{g}_{\mu\nu}(q)$. A tensor transformation
of the (contravariant) metric is $\bar{g}^{\rho\sigma}(q)=\partial_{\mu
}q^{\rho}\partial_{\nu}q^{\sigma}g^{\mu\nu}(x)$. The constraint equations
$\partial_{\mu}q^{\alpha}\partial^{\mu}q^{\beta}=\eta^{\alpha\beta}$ state
that the $\alpha\beta$ components of $\bar{g}_{\rho\sigma}(q)$ -- a subset of
the full set of $\{\bar{g}_{\rho\sigma}\}$ -- become Minkowski-valued on the
solution manifold $\mathcal{M}$. The solution $\phi(x^{\mu})$ to the EoM is
mapped into a function $\phi(q^{\alpha})$, which solves a DE (on $\mathcal{M}%
$) with fewer (generalized) coordinate variables on a manifold $\mathcal{M}$
where some of the metric components $\bar{g}_{\rho\sigma}$ take Minkowski values.

\section{Some illustrations}

A few concrete illustrations are given for implementing the method described
above. We focus on cases where there are only one or two $q$ functions, i.e.,
the $q^{\alpha}-$ space (the number of $q$'s on which $\phi$ depends) is one
or two dimensional.

\subsection{The 1d case}

\textbf{Spacelike case:}\ \ Let us seek a solution to the EoM involving one
spacelike function, say $q^{1}=\xi(x^{\mu})$ so that the solution to the EoM
$\square\phi(x^{\mu})+V^{\prime}(\phi)=0$ on the manifold $\mathcal{M}$ is
given by $\phi\lbrack\xi(x^{\mu})]$. The function $\xi$ must be harmonic,
$\nabla_{\mu}\partial^{\mu}\xi=\square\xi=0$, and must satisfy the constraint
in (\ref{7a}) which takes the form%
\begin{equation}
\partial_{\mu}\xi\partial^{\mu}\xi=-1 \label{9}%
\end{equation}

Nonlinear harmonic functions $\xi$ will solve this constraint when the
coordinates $x^{\mu}$ are constrained, and thereby define a manifold
$\mathcal{M}$. For example, consider the spacetime to be a 4d Minkowski
spacetime, $g_{\mu\nu}(x)=\eta_{\mu\nu}$, and choose the harmonic function
$\xi_{R}=xy/R$, where $R$ is an arbitrary real, positive constant. The
constraint (\ref{9}) then becomes the condition%
\begin{equation}
x^{2}+y^{2}=R^{2} \label{10}%
\end{equation}

so that the spatial surface $\mathcal{M}_{R}$ \ is a static cylinder of radius
$R$ centered on the $z$ axis. Then the solution to the EoM on $\mathcal{M}%
_{R}$, where $\xi_{R}=xy/R=R\sin\theta\cos\theta$ (with $\theta$ the ordinary
azimuth angle) is $\phi_{R}(\theta)=\phi_{R}(R\sin\theta\cos\theta)$. Since
$R$ is a continuous real parameter, there is a continuum of surfaces
$\mathcal{M}_{R}$ (concentric cylinders) on which solutions $\phi_{R}$ to the
EoM exist. The space is then foliated by a set of concentric cylindrical
leaves, with a solution $\phi_{R}(\theta)$ defined on each leaf labelled by
the parameter $R$. Looking at the leaf solutions as $R$ ranges from zero to
infinity can give a glimpse of qualitative features expected of an exact
solution $\Phi(x^{\mu})$ to the EoM\ $\square\Phi+V^{\prime}(\Phi)=0$ that
exists in the bulk of the spacetime, i.e., a solution that satisfies the EoM
throughout the entire spacetime. (Each of these leaf solutions $\phi_{R}$
generally has a nonvanishing normal derivative $\hat{n}\cdot\nabla\phi$ on the
surface $\mathcal{M}_{R}$ in addition to tangential derivatives along the
surface. The solution $\phi(\xi_{R})$ takes a value of $\phi_{R}(\xi_{R}%
)=\phi(\xi_{R})|_{\mathcal{M}_{R}}$ on the surface $\mathcal{M}_{R}$ where
$\xi_{R}$ takes a value $\xi_{R}=\left(  r^{2}/R\right)  \sin\theta\cos
\theta|_{r=R}=R\sin\theta\cos\theta$.)

This 1d case is an illustration of a \textquotedblleft
BPS-like\textquotedblright\ ansatz, since the simplified equation in (\ref{6})
is just%
\begin{equation}
-\partial_{\xi}^{2}\phi(\xi)+V^{\prime}(\phi)=0 \label{11}%
\end{equation}

and can be integrated to give%
\begin{equation}
\tfrac{1}{\sqrt{2}}\partial_{\xi}\phi=\pm\sqrt{V+c} \label{12}%
\end{equation}

where $c$ is an integration constant, determined by boundary conditions. The
solution is then given by%
\begin{equation}
\int\frac{d\phi}{\sqrt{V+c}}=\pm\sqrt{2}(\xi-\xi_{0}) \label{13}%
\end{equation}

which can be determined explicitly, once the form of the potential $V(\phi)$
is specified. The 2nd order EoM has been transformed into the 1st order DE in
(\ref{12}), which resembles the DE for a BPS solution for a static field which
is a function of the coordinate $\xi$. This BPS-like ansatz can be used to
obtain new solutions on various manifolds for different scalar field theories.
Specific examples follow. (We assume a 4d Minkowski spacetime.)

\medskip

(1) \textit{Lorentz boosted kink}:\ \ For a specific example, consider
$\phi^{4}$ theory with potential $V=(\phi^{2}-1)^{2}$. Choosing $c=0$,
(\ref{13}) gives the familiar kink solution $\phi(\xi)=\tanh(\sqrt{2}\ \xi)$.
Let us now choose a linear harmonic function, $\xi=a_{\mu}x^{\mu}$. The
constraint (\ref{9}) leads to $a_{\mu}a^{\mu}=-1$, which does not involve
coordinates, but only constrains the constants $a_{\mu}$. Therefore the
solution manifold $\mathcal{M}$ is the full spacetime. Note that this choice
of $\xi$ includes a description of a Lorentz boost, as can be seen by choosing
$a_{0}=-\gamma u$, $a_{1}=\gamma$, $a_{2}=a_{3}=0$. The constraint has as a
solution $\gamma=(1-u^{2})^{-1/2}$, which is the relativistic $\gamma$ factor
associated with a boost along the $x$ axis with velocity $u$. Then $\xi
=\gamma(x-ut)$ gives a Lorentz transform from $x$ to $x^{\prime}=\xi(x,t)$.
The kink solution $\phi(\xi)$ therefore can be written as $\phi(x,t)=\tanh
[\sqrt{2}\gamma(x-ut)]$, a Lorentz boosted kink defined in the whole
spacetime. (Linear functions $q^{\alpha}$ in a Minkowski spacetime generate
constraints involving only constants, rather than coordinates. Nonlinear
functions $q^{\alpha}$ are associated with coordinate-constrained manifolds.)

\medskip

(2) $\phi^{4}$ \textit{domain ribbons on static cylinder}:\ \ As another
example, consider $\phi^{4}$ kink solutions on the surface of the cylinder of
radius $R$ in (\ref{10}), generated by the function $\xi_{R}=xy/R=(r^{2}%
/R)\sin\theta\cos\theta$. On the surface $\mathcal{M}_{R}$ this takes the
value $\xi_{R}|_{\mathcal{M}}=R\sin\theta\cos\theta$. The kink solutions
$\phi(\xi)=\pm\tanh(\sqrt{2}\ \xi)$ on the cylinder surface $\mathcal{M}_{R}$ are%

\begin{equation}
\phi_{R}(\xi_{R})=\pm\tanh\left(  \sqrt{2}R\sin\theta\cos\theta\right)
\label{a1}%
\end{equation}

These are $z$ independent solutions with zeros located on the $\pm x$ and $\pm
y$ axes. The energy density is%
\begin{equation}
T_{00}=g_{00}[2V]\medskip=\frac{2}{\cosh^{4}(\sqrt{2}R\sin\theta\cos\theta)}
\label{a2}%
\end{equation}

This energy density is maximized at the zeros of the solution $\phi$; we can
think of these solutions as domain ribbons on the cylinder, parallel to the
$z$ axis. For either the ($+$) or ($-$) solutions, we have zeros of $\phi$
with positive slopes separated by zeros of $\phi$ with negative slopes in
between. This leads us to interpret the solution as a set of four ribbon-like
structures consisting of two ribbons separated by antiribbons in between.

As the parameter $R$ ranges from zero to infinity, we infer from the
$\{\phi_{R}(\xi_{R})\}$ the existence of a static bulk solution $\Phi(x,y)$
describing perpendicular domain walls centered on the $x$ and $y$ axes, where
$\Phi=0$, with $\Phi$ entering vacuum states $\Phi=\pm1$ away from the axes at
asymptotic distances from the origin. The set of surface solutions $\left\{
\phi_{R}\right\}  $ presumably resemble intersections of a bulk solution
$\Phi$ with the leaves of the $\{\mathcal{M}_{R}\}$ surfaces.

\medskip

\textbf{Timelike case:}\ \ If we instead consider a single timelike
generalized coordinate $\tau(x^{\mu})$, the EoM reduces to $\partial_{\tau
}^{2}\phi(\tau)+V^{\prime}(\phi)=0$ with the harmonic function $\tau$ subject
to the constraint $\partial_{\mu}\tau\partial^{\mu}\tau=\eta^{00}=1$. The DE
for $\phi(\tau)$ can be solved once the form of the potential (along with
boundary conditions) is specified. The manifold $\mathcal{M}$ is generated by
the choice of $\tau$ and the constraint that it must satisfy.

\medskip

\textit{K-G field on dynamical 2-branes}:\ As an example, in a 4d Minkowski
spacetime, a potential $V=\frac{1}{2}m^{2}\phi^{2}$ admits a simple solution
$\phi(\tau)=\cos m\tau$. Choosing, for example, a function $\tau=xt$ leads to
a constraint $x^{2}-t^{2}=1$, which defines two parallel planes perpendicular
to the $x$ axis, located by
\begin{equation}
x^{\pm}(t)=\pm\sqrt{t^{2}+1} \label{14}%
\end{equation}

The planes approach one another for $t<0$, stop and turn around at $t=0$, then
move away from each other for $t>0$. The value of $\tau^{\pm}$ on
$\mathcal{M}^{\pm}$ is $\tau^{\pm}=x^{\pm}t=\pm t\sqrt{t^{2}-1}=\pm x^{\pm
}\sqrt{(x^{\pm})^{2}-1}$. The solution $\phi(x,t)$ of the EoM can then be
written, for instance, as
\begin{equation}
\phi(x,t)=\cos m\tau=\cos(mxt) \label{15}%
\end{equation}

This function satisfies the EoM $(\partial_{t}^{2}-\partial_{x}^{2}%
)\phi(x,t)+m^{2}\phi(x,t)=0$ \textit{when the EoM is evaluated on the
manifold} $\mathcal{M}$. The value of the solution $\phi(x^{\pm}t)$ on
$\mathcal{M}^{\pm}$ is then given by%
\begin{equation}
\phi_{\mathcal{M}^{\pm}}(t)=\cos m\tau^{\pm}=\cos\left[  mt\sqrt{t^{2}%
-1}\right]  \label{16}%
\end{equation}

Keep in mind that it is not (\ref{16}) that solves the EoM on $\mathcal{M}$,
but rather the function in (\ref{15}), which has nonvanishing normal
derivatives ( $x$-derivatives). The solution of (\ref{15}) then takes the
value given by (\ref{16}) on the surfaces $\mathcal{M}^{\pm}$ where $x=x^{\pm
}$.

\subsection{The 2d case}

\textbf{1+1 case:\ \ }Consider $\phi$ to be a function of just two $q$'s, say
a timelike function $q^{0}=\tau(x^{\mu})$ and a spacelike function $q^{1}%
=\xi(x^{\mu})$, so that $\phi=\phi(\tau,\xi)$. Then the conditions in
(\ref{7a}) are given explicitly by the harmonic conditions $\square
\tau=\square\xi=0$ supplemented by the set of constraints%

\begin{equation}%
\begin{array}
[c]{lll}%
\partial_{\mu}q^{0}\partial^{\mu}q^{0}=\eta^{00} &  & \partial_{\mu}%
\tau\partial^{\mu}\tau=1\\
\partial_{\mu}q^{0}\partial^{\mu}q^{1}=\eta^{01} & \ \ \ \text{or\ \ \ } &
\partial_{\mu}\tau\partial^{\mu}\xi=0\\
\partial_{\mu}q^{1}\partial^{\mu}q^{1}=\eta^{11} &  & \partial_{\mu}%
\xi\partial^{\mu}\xi=-1
\end{array}
\label{17}%
\end{equation}

This set of simultaneous constraints can, in general, lead to intersecting
surfaces, etc., and the solution manifold, $\mathcal{M}$, is the common
intersection of all the individual constraint manifolds. The scalar field
$\phi\lbrack\tau(x^{\mu}),\xi(x^{\mu})]$ is a solution of the simplified EoM%
\begin{equation}
(\partial_{\tau}^{2}-\partial_{\xi}^{2})\phi+V^{\prime}(\phi)=0 \label{18}%
\end{equation}

and this solution solves the original EoM $\nabla_{\mu}\partial^{\mu}%
\phi+V^{\prime}(\phi)=0$ on the solution manifold $\mathcal{M}$. We give
specific examples below. (We assume a flat 4d spacetime.)

\medskip

(1)\ \textit{Massless scalar field}:\ \ For a potential $V(\phi)=0$ the
general solution of (\ref{18}) is%
\begin{equation}
\phi(\tau,\xi)=F(\tau+\xi)+G(\tau-\xi) \label{19}%
\end{equation}

where $F$ and $G$ are arbitrary functions of the indicated arguments and
$\tau(x^{\mu})$ and $\xi(x^{\mu})$ are functions that satisfy (\ref{17}). An
example of such $\tau$ and $\xi$ functions is%
\begin{equation}
\tau=\sqrt{2}t-z,\ \ \ \xi=xy=r^{2}\sin\theta\cos\theta\label{20}%
\end{equation}

for which $\mathcal{M}$ is a static cylinder of unit radius centered on the
$z$ axis. Then on the cylindrical surface $\mathcal{M}$ the solution in
(\ref{19}) takes the form
\begin{equation}
\phi(\tau,\xi)\Big|_{\mathcal{M}}=\phi_{\mathcal{M}}(t,z,\theta)=F(\sqrt
{2}t-z+\sin\theta\cos\theta)+G(\sqrt{2}t-z-\sin\theta\cos\theta) \label{21}%
\end{equation}

These running waves have the form $f(\sqrt{2}t-\zeta_{\pm})$, with $\zeta
_{\pm}=z\pm\sin\theta\cos\theta$.

\medskip

(2) \textit{Massive Klein-Gordon field}:\ \ \ For a potential $V(\phi
)=\frac{1}{2}m^{2}\phi^{2}$ a simple wavelike solution of (\ref{18}) is%
\begin{equation}
\phi=\cos(\omega\tau-k\xi),\ \ \ \omega^{2}=k^{2}+m^{2} \label{22}%
\end{equation}

We choose the same manifold functions as before, given in (\ref{20}). The
ansatz solution is then%
\begin{equation}
\phi=\cos\left[  \omega\left(  \sqrt{2}t-z\right)  -kr^{2}\sin\theta\cos
\theta\right]  \label{23}%
\end{equation}

and on the cylinder $\mathcal{M}$ we set $r=1$. We could write this as
$\phi_{\mathcal{M}}=\cos\left[  \Omega\ t-Kz+\delta(\theta)\right]  $, with
$\Omega=\sqrt{2}\omega$, $K=\omega$, and phase parameter $\delta
(\theta)=-k\sin\theta\cos\theta$. The condition $\omega^{2}-k^{2}=m^{2}$
gives
\begin{equation}
\Omega^{2}-K^{2}=\omega^{2}=k^{2}+m^{2}\equiv M^{2} \label{24}%
\end{equation}

So (\ref{23}) and (\ref{24}) describe a massive plane wave traveling in the
$z$ direction on the cylinder, with energy $\Omega$, momentum $K$, and
effective mass $M=\sqrt{k^{2}+m^{2}}$. There is an angular dependent phase
constant $xy=\sin\theta\cos\theta$ which vanishes on the $x$ and $y$ axes, but
becomes nonzero elsewhere.

\medskip

(3) \textit{Dynamical }$\phi^{4}$\textit{ domain ribbons}:\ \ \ For a
potential $V(\phi)=(\phi^{2}-1)^{2}$ we displayed a static solution for a kink
as $\phi(\xi)=\tanh(\sqrt{2}\ \xi)$ for the 1d case above. For a simple 2d
solution satisfying (\ref{18}) we take a Lorentz boosted version of $\phi
(\xi)$, with $\xi\rightarrow\gamma(\xi-u\tau)$, which we write as%
\begin{equation}
\phi(\tau,\xi)=\tanh\left[  \sqrt{2}\ \gamma(\xi-u\tau)\right]  \label{25}%
\end{equation}

We again choose the functions $\tau$ and $\xi$ in (\ref{20}). The ansatz
solution on the cylinder then takes the form%
\begin{equation}
\phi(\tau,\xi)\Big|_{\mathcal{M}}=\tanh\left\{  \sqrt{2}\ \gamma\left[
\sin\theta\cos\theta-u(\sqrt{2}t-z)\right]  \right\}  \label{26}%
\end{equation}

For $u=0$, $\gamma=1$ this describes a pair of domain ribbons, each ribbon
separated from the next by an antiribbon, all lying parallel to the $z$ axis
and centered on the $\pm x$ and $\pm y$ axes, where the energy density
maximizes (at $\phi=0$, or $xy=0$). However, for $u\neq0$ the zeros of $\phi$
are shifted to positions located by $xy=\sin\theta\cos\theta=u(\sqrt{2}t-z)$,
indicating that the locations of the ribbon cores on the cylinder wall become
$z$ and $t$ dependent dynamical objects. For instance, at the time $t=0$ we
have ribbons localized at $xy=\sin\theta\cos\theta=-uz$ so that the ribbons
appear to wind around the cylinder in a helical fashion, and these windings
move as $t$ progresses.

\medskip

\textbf{2+0 case:}\ \ Now consider a type of solution where $\phi$ depends on
two spacelike generalized coordinates $q^{1}=\xi(x^{\mu})$ and $q^{2}%
=\sigma(x^{\mu})$. The equation of motion in (\ref{6}) becomes%
\begin{equation}
(\partial_{\xi}^{2}+\partial_{\sigma}^{2})\phi(\xi,\sigma)=V^{\prime}(\phi)
\label{27}%
\end{equation}

with $\square\xi=\square\sigma=0$. The constraints in (\ref{7a}) take the form%
\begin{equation}%
\begin{array}
[c]{lll}%
\partial_{\mu}q^{1}\partial^{\mu}q^{1}=\eta^{11} &  & \partial_{\mu}%
\xi\partial^{\mu}\xi=-1\\
\partial_{\mu}q^{1}\partial^{\mu}q^{2}=\eta^{12} & \ \ \ \text{or\ \ \ } &
\partial_{\mu}\xi\partial^{\mu}\sigma=0\\
\partial_{\mu}q^{2}\partial^{\mu}q^{2}=\eta^{22} &  & \partial_{\mu}%
\sigma\partial^{\mu}\sigma=-1
\end{array}
\label{28}%
\end{equation}

\medskip

\textit{Laplace's equation on a cylinder}:\ \ Example constraint functions are%
\begin{equation}
\xi=xy=r^{2}\sin\theta\cos\theta,\ \ \ \ \sigma=\gamma(z-ut),\ \ \gamma
=1/\sqrt{1-u^{2}} \label{29}%
\end{equation}

which describe Lorentz boosts in the $z$ direction on the surface of a
cylinder of unit radius, centered on the $z$ axis. As an example of a
potential, we choose that of a massless scalar field, $V(\phi)=0$. In this
case a general solution to (\ref{27}) can be written as%
\begin{equation}
\phi(\xi,\sigma)=\sum_{k}A_{k}e^{-k\xi}\cos k\sigma\label{30}%
\end{equation}

For the $\xi$ and $\sigma$ chosen above, the solution on the cylinder becomes%
\begin{equation}
\phi_{\mathcal{M}}=\sum_{k}A_{k}e^{-k\sin\theta\cos\theta}\cos k\gamma(z-ut)
\label{31}%
\end{equation}

Each $k$ solution varies in a periodic way around the cylinder in the $\theta$
direction, and is also a periodic function of $z-ut$. The values of $k$ and
the constants $A_{k}$ are determined by boundary conditions.

\section{Summary}

A method has been presented which allows a class of nontrivial solutions to
the equation of motion (EoM) for a real scalar field $\phi(x^{\mu})$, given by
$\square\phi+V^{\prime}(\phi)=0$, to be obtained from a simplified equation of
motion. This is accomplished by replacing coordinate variables $x^{\alpha}$ on
which a scalar field $\varphi$ depends with generalized coordinates
$q^{\alpha}(x^{\mu})$, which are harmonic functions of coordinates $x^{\mu}$.
The function $\varphi(x^{\alpha})$ satisfies the simpler equation
$\eta^{\alpha\beta}\partial_{\alpha}\partial_{\beta}\varphi(x)+V^{\prime
}(\varphi)=0$, with the $\{x^{\alpha}\}$ being a subset of the full set of
coordinates $\{x^{\mu}\}$. The replacements $x^{\alpha}\rightarrow q^{\alpha}$
and $\varphi(x^{\alpha})\rightarrow\phi(q^{\alpha})$ results in a function
$\phi(x^{\mu})=\phi\lbrack q^{\alpha}(x^{\mu})]$ that solves the original EoM
$\nabla_{\mu}\partial^{\mu}\phi(x)+V^{\prime}(\phi)=0$, provided that a set of
simple constraints is satisfied. These constraints give rise to spacetime
manifolds $\mathcal{M}$ on which the solution $\phi(x^{\mu})$ exists. In a
Minkowski spacetime, linear functions $q^{\alpha}(x^{\mu})$ are associated
with a manifold which is the full spacetime, with constraints on the
constants, whereas for nonlinear functions $q^{\alpha}(x^{\mu})$ the manifold
is a subspace or hypersurface of the spacetime. Neither the manifolds nor
their dynamics depend upon the form of the scalar field theory. Examples of
manifolds and solutions for different scalar field theories have been provided
for the 1d and 2d cases, i.e., where the function $\phi$ depends on only one
or two generalized coordinate functions $q^{\alpha}$. Dynamical manifolds, or
a continuum of static manifolds, can span the bulk of the spacetime, allowing
some inference of the nature of exact bulk solutions $\Phi(x^{\mu})$ that
solve the EoM throughout the entire spacetime, without being restricted to any
particular manifold.

\end{document}